\documentclass[a4paper,fleqn]{cas-dc}
\usepackage[numbers,sort&compress]{natbib}
\usepackage{xcolor}

\def\tsc#1{\csdef{#1}{\textsc{\lowercase{#1}}\xspace}}
\tsc{WGM}
\tsc{QE}
\tsc{EP}
\tsc{PMS}
\tsc{BEC}
\tsc{DE}

\ExplSyntaxOn
\cs_set:Npn \__first_footerline: {}
\ExplSyntaxOff

\begin{document}
\let\WriteBookmarks\relax
\def\floatpagepagefraction{1}
\def\textpagefraction{.001}

\shorttitle{Lanthanide Doped YSO Microcrystals for QIP}

\shortauthors{J.L.B, Martin et~al.}

\title [mode = title]{Growth and Spectroscopy of Lanthanide Doped Y$_2$SiO$_5$ Microcrystals for Quantum Information Processing}

\author[1,2]{Jamin L. B. Martin}[orcid=0000-0002-7204-231X]

\credit{Conceptualisation, Methodology, Formal analysis, Investigation, Visualisation, Writing - Original Draft}

\author[1]{Lily F. Williams}
\credit{Investigation,Visualisation}

\address[1]{The School of Physical and Chemical Sciences, University of Canterbury, PB4800 Christchurch 8140, New Zealand}
\author[1,2]{Michael F. Reid}[orcid=0000-0002-2984-9951]
\cormark[1]
\credit{Supervision, Writing Review \& Editing, Conceptualisation, Resources}
\ead{mike.reid@canterbury.ac.nz}
\author%
[1,2]{Jon-Paul R. Wells}[orcid=0000-0002-8421-6604]
\cormark[1]
\credit{Supervision, Writing Review \& Editing, Conceptualisation, Resources}
\ead{jon-paul.wells@canterbury.ac.nz}

\address[2]{The Dodd-Walls Centre for Photonic and Quantum Technologies, New Zealand}

\cortext[cor1]{Corresponding authors}

\begin{abstract}
Lanthanide-doped Y$_{2}$SiO$_{5}$ microcrystals were prepared using the solution combustion, solid state and sol-gel synthesis techniques. Of these, the sol-gel method yields the most reliable and high-quality X2 phase Y$_{2}$SiO$_{5}$ microcrystals. Absorption and laser site-selective fluorescence measurements of Nd$^{3+}$, Eu$^{3+}$ and Er$^{3+}$ doped material, performed at cryogenic temperatures, indicate that the as-grown microcrystals are of high optical quality with inhomogeneously broadened optical linewidths that are comparable to bulk crystals at similar dopant concentrations.
\end{abstract}



\begin{keywords}
Microcrystals \sep Erbium \sep Neodymium \sep Europium \sep Spectroscopy
\end{keywords}

\maketitle

\section{Introduction}
Modern quantum technologies are in a state of rapid development. These range from well-proven systems such as superconducting qubits \cite{clarke2008superconducting,arute2019quantum} and artificially trapped ions \cite{bruzewicz2019trapped,kaushal2020shuttling} to emerging platforms such as NV centres in diamond \cite{pezzagna2021quantum,nakazato2022quantum} and the use of rare-earth, in particular lanthanide ions. Rare-earth ions offer a myriad of benefits including excellent coherence properties \cite{konz2003temperature} and broad inhomogeneous linewidths to facilitate a large number of qubit channels. More so, the large variety of crystalline host materials, as well as possible dopant ions, create a significant matrix of material combinations to meet a given technological requirement, e.g. telecoms integration utilising the 1.5\,$\mu$m erbium transition. 

Given the potential of lanthanide-doped insulating dielectric materials, work has been carried out to integrate them into smaller devices. Most commonly, host crystals are grown in a high-temperature melt and are macroscopic (i.e. millimetres to centimetres) in size. This presents issues around miniaturisation, with complex etching or machining required \cite{zhong2017nanophotonic,zhong2016high,yang2021photonic}. An alternative approach is the growth of nano/micro-crystals for integration into fibre-based scanning Fabry-Perot microcavities \cite{casabone2018cavity}, the growth of thin films through chemical vapour deposition \cite{flinn1994anomalous} and atomic layer deposition techniques \cite{scarafagio2019ultrathin}. There is also evidence that lanthanide-doped microcrystals have the potential to be utilised for CNOT and other conditional gate operations in optical quantum computers, with high speed and fidelity owing to their size \cite{hizhnyakov2021rare}. Despite the promise of miniaturisation and novel application approaches, the growth of nano/micro-crystals is also fraught with limitations. Nano/micro-crystals prepared using different growth methods often have quite different optical characteristics. For example, nano/micro-crystals grown by simple hydrothermal techniques offer great size control however they suffer from the inclusion of -OH groups that act as strong quenchers of luminescence \cite{samsonova2016fluorescence} or surface states which can similarly modify luminescence properties. In some cases, the growth process can inhibit the formation of optical centres present in the bulk crystal, complicating the miniaturisation process \cite{balabhadra2020absorption}. Any bulk to nano variation in the electronic structure of an optically active ion needs to be accounted for, if established techniques such as crystal-field calculations are to be used to predict Zeeman splittings and/or optical properties such as transition intensities or polarisation behaviour \cite{MARTIN2022100181}. Such calculations are valuable in the development of quantum memories where estimation of Zeeman-hyperfine interactions are highly relevant \cite{jobbitt2022zeeman,jobbitt2021prediction}. 

One material where there has been considerable recent research in the quantum technology context is Y$_2$SiO$_5$ (YSO). Hyperfine coherence times of 1 minute and 6 hours have been demonstrated in bulk crystals of YSO:Pr$^{3+}$ and YSO:Eu$^{3+}$ respectively \cite{heinze2013stopped,zhong2015optically}, at so-called ZEFOZ (zero first order Zeeman effect) points. YSO is an ideal host for quantum information storage applications due to the low abundance of nuclear spins in the material which acts to minimise fluctuations in the local magnetic field environment. Yttrium has a nuclear spin of 1/2 while silicon and oxygen have few abundant isotopes with non-zero nuclear spins. YSO has also been used in the development of quantum gates and single photon sources \cite{dibos2018atomic,de2008solid,rippe2008experimental}. YSO is a monoclinic crystal, with its low and high-temperature phases offering two yttrium sites for lanthanide ion substitution, with both having C$_1$ point group symmetry. Bulk crystals prepared using the Czochralski technique only form in the X2 phase, which has the space group C2/c \cite{maksimov1970crystal}. Micro- or nano-crystals formed at temperatures below $\sim$1200 degrees crystallise in the X1 phase with the space group P21/c \cite{cooke2006luminescent}. 
There are many reports of preparation of micro- and nano-crystalline Y$_2$SiO$_5$. However, these studies tend to be oriented toward applications such as \cite{ghosh2006preparation,dramicanin2008synthesis,zhang2002luminescent} the development of scintillators \cite{popovich2019highly}, WLED phosphors \cite{ramakrishna2016white}, solar converters \cite{kang2018enhanced}, radiation dosimetry \cite{ramakrishna2014effect}, and upconverting materials for biomedical imaging \cite{sengthong2016bright}. The yttrium-silicate family form a complex system of possible materials depending on the exact stoichiometry and synthesis temperature \cite{liu2022spherical}. Therefore, it is common to find a degree (sometimes substantial) of phase impurity upon close inspection of crystallographic data. This is highly problematic if you require high-quality, high-resolution spectroscopic data to inform calculations of electronic structure which in turn are able to predict (for example) ZEFOZ points. Therefore, finding a preparation technique that is both scalable for technological applications and high enough quality for fundamental spectroscopy is paramount. 

In this work, we present the synthesis, structural and spectroscopic assay of lanthanide-doped Y$_2$SiO$_5$ microcrystals. We report a synthesis method that produces microcrystals with the same structure, and comparable spectroscopic properties, to bulk crystals. This paves the way for a higher throughput analysis of lanthanide-doped Y$_2$SiO$_5$ to enhance current efforts in modelling electronic structure as well as the potential integration into micro-cavities, low-cost targets for pulsed laser deposited thin films, and microcrystal-based optical quantum computers.

\section{Methods}

\subsection{Materials}
Ytterbium nitrate (Yb(NO$_3$)$_{3} \cdot 6$H$_{2}$O, 99.999\%), \newline praseodymium nitrate (Pr(NO$_3$)$_{3} \cdot 5$H$_{2}$O, 99.9\%), europium nitrate (Eu(NO$_3$)$_{3} \cdot x$H$_{2}$O, 99.99\%) , holmium nitrate (Ho(NO$_3$)$_{3} \cdot 6$H$_{2}$O, 99.9\%), erbium nitrate (Er(NO$_3$)$_{3} \cdot 6$H$_{2}$O, 99.9\%) (Sigma-Alrdrich, St. Louis, USA), yttrium nitrate (Y(NO$_3$)$_{3} \cdot 6$H$_{2}$O, 99.9\%) (Alfa Aesar, Haverhill, Massachusetts, USA), neodymium nitrate (Nd(NO$_3$)$_{3} \cdot x$H$_{2}$O, 99.9\%), cerium nitrate (Ce(NO$_3$)$_{3} \cdot x$H$_{2}$O, 99.9\%), europium oxide (Eu$_2$O$_3$, 99.9\%), yttrium oxide (Y$_2$O$_3$, 99.9\%) (AMPOT, USA), TEOS (C$_8$H$_{20}$O$_{4}$Si) (Alfa Aesar, Haverhill, Massachusetts, USA) Silica (SiO$_2$, 99.9999\%) (Sigma-Aldrich, St. Louis, USA), ammonia solution 35\% (NH$_4$OH) (Waltham, Massachusetts, United States). 

\subsection{Synthesis}

Preparation of X2 phase, YSO nano-/microcrystals was attempted using three different methods. These include solution combustion, solid-state and sol-gel synthesis. Of these, only the sol-gel synthesis method was found to produce reliable, high-quality X2-YSO micro-crystals. The solution combustion and solid state synthesis methods were performed as given in refs \cite{ramakrishna2014effect,ramakrishna2016white,rakov2014cooperative,chepyga2018synthesis} with the volumes of precursors scaled to that of what is used in for the sol-gel method, to maintain a like for like comparison.

\begin{figure}[h!]
	\centering
	\includegraphics[width=\linewidth]{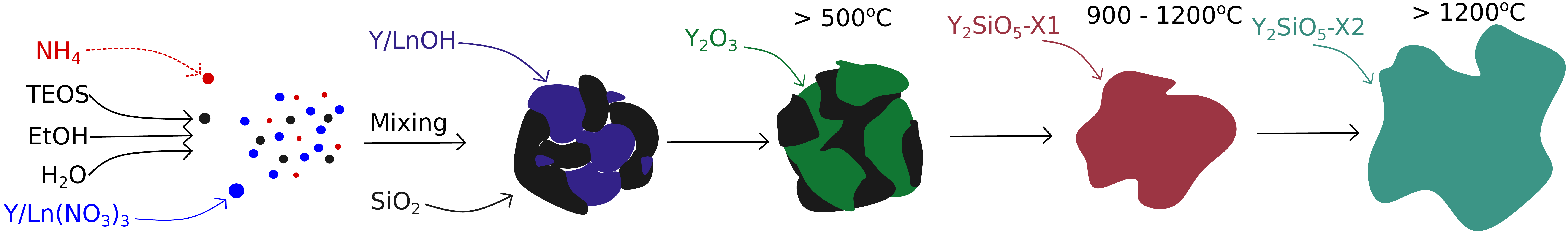}
	\caption{Schematic of the growth process of YSO microcrystals.}
	\label{ysogrowth}
\end{figure}

X2 phase YSO micro-crystals were successfully and reliably grown via a sol-gel route similar to ref \cite{popovich2019highly}. The method follows a modified St{\"o}ber process for producing spherical silica nanoparticles \cite{stober1968controlled}; in most cases, ammonia is added as a catalyst to speed up the condensation-hydration reaction to form silica spheres. Here, the ammonia acts as both a catalyst and a reactant for the formation of lanthanide/yttrium hydroxides. In brief, 0.98\,mmol of Y(NO$_3$)$_{3} \cdot 6$H$_{2}$O and 0.02\,mmol of Ln(NO$_3$)$_{3} \cdot x$H$_{2}$O (lanthanide hydration state varies across the series) is mixed vigorously in 10ml of deionised water (solution one). Solution two is prepared by vigorously mixing 0.5\,mmol of TEOS with 10\,ml of ethanol. Solution two is mixed with solution one under vigorous stirring until the solution becomes clear. Lastly, 2\,ml of 30\% ammonia water is added dropwise under vigorous stirring. The solution was left to mix for 24\,h. The solution was centrifuged and washed five times, alternating between water and ethanol. The washed precipitate was dried for 48\,h at 50$^{\circ}$C and then sintered in an alumina crucible at 1400$^\circ$C for 4 hours, with a heating and cooling rate of 5 degrees per minute in a SentroTech ST-1800 muffle furnace. A schematic of the growth process is shown in fig \ref{ysogrowth}. A sintering temperature of 1400$^\circ$C was used as it is well above the 1200-1300$^\circ$C temperature range where X1-YSO is believed to undergo a phase transition to X2-YSO. 

\subsection{Characterisation}

Phase identification of the YSO samples was inferred from their X-ray diffraction patterns. Diffraction patterns were collected on a RIGAKU 3kW SmartLab diffractometer equipped with a Cu K$\alpha$ radiation source ($\lambda$ = 1.5406 \AA{}), operating at 40 kV and 30 mA, with a 0.01$^{\circ}$ step size in reflection scanning mode. The observed diffraction patterns were compared to reference data taken from the International Centre for Diffraction Data (ICDD) database.

Size and morphology were observed via scanning electron microscopy (SEM). Images were recorded using a JEOL JSM 7000F field emission, high-resolution scanning electron microscope, operating at 15\,kV with a 7-8mm working distance. Samples were prepared by sonicating the prepared crystals in acetone followed by dispersing them onto a carbon conductive tab. 

Infrared absorption spectroscopy was performed using an N$_{2}$ gas-purged Bruker Vertex 80 FTIR equipped with liquid nitrogen-cooled mercury cadmium telluride, thermo-electrically cooled indium gallium arsenide, and room temperature silicon detectors giving a spectral range of 350-25,000 cm$^{-1}$. A tungsten-halogen lamp and Globar were used as a NIR and MIR sources, respectively. Samples were cooled to 10\, K using a Janis closed cycle cryostat in a transmission configuration. Spectra were recorded with a resolution of 0.2\,cm$^{-1}$ (microcrystals) and 0.075\,cm$^{-1}$ (bulk crystals).

Laser site selective spectroscopy was performed using microcrystal samples pressed into the recess of a copper cold-finger, which was capped with a CaF$_2$ cover. Samples were cooled to 10\,K using a Janis closed-cycle cryostat. The samples were excited using a PTI tunable dye laser (using both Coumarin 540A and Coumarin 460 dyes ), which was pumped by a PTI pulsed nitrogen laser at a 5\,Hz rep rate. Spectra were recorded using an iHR550 spectrometer (Horiba Scientific) with a 1200 grooves$\cdot$mm\textsuperscript{-1} grating coupled to a Hamamatsu R2257P thermoelectrically cooled PMT.

\section{Results}

\subsection{Morphology and Crystalinity}
\label{xrd_sem}
\begin{figure}[ht!]
	\centering
	\includegraphics[width=\linewidth]{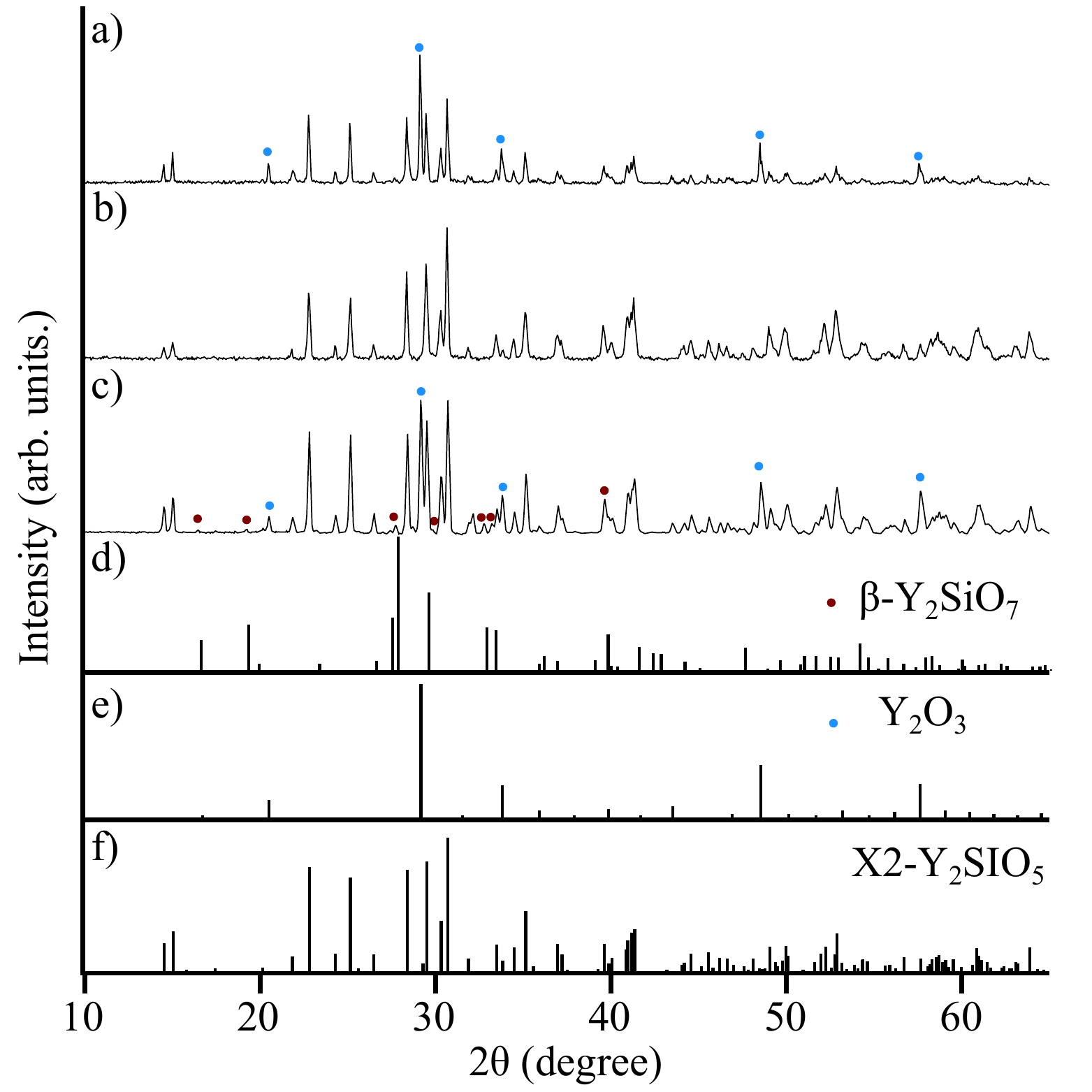}
	\caption{X-ray diffractograms of lanthanide-doped YSO micro-crystals produced by differing methods a) solid state synthesis b) sol-gel synthesis, c) solution combustion synthesis, d) $\beta$-Y$_2$SiO$_7$ reference ICCD card No. 04-012-4410, e) Y$_2$O$_3$  reference ICCD card No. 00-041-1105 and X2-YSO reference ICCD card No. 01-084-4282 }
	\label{YSO_method}
\end{figure}

Figure \ref{YSO_method} presents representative X-ray diffractograms for samples prepared using the three different synthesis types investigated. All samples show a dominant amount of X2-YSO has formed, with no indication of the X1 phase being present. Most notably, all but the sol-gel method (Fig \ref{YSO_method} b)) have characteristic peaks of $\beta$-Y$_2$Si$_2$O$_7$ and Y$_2$O$_3$, as indicated by the reference data. Whilst this could be explained by a miss measurement of the yttrium or silica precursors, however, these samples were prepared multiple times with similar outcomes, with the sol-gel method producing higher-quality crystals as indicated by the diffraction patterns. In the case of the solution combustion method (Fig \ref{YSO_method} c)), both Y$_2$O$_3$ and $\beta$-Y$_2$Si$_2$O$_7$ are present. The presence of the Y$_2$O$_3$ and Y$_2$Si$_2$O$_7$ phases in samples prepared by solution combustion is documented \cite{rakov2014cooperative}, indicating it is not stoichiometry alone driving the phase impurities but more likely the method itself in combination with the precursor materials, additionally the observed ratio of Y$_2$Si$_2$O$_7$:Y$_2$O$_3$ seems to vary, both by our observation and inspection of the literature. Concerning the solid-state synthesis method (Fig \ref{YSO_method} a)), the presence of Y$_2$O$_3$ is likely due to the combination of temperature and relative size of the precursors. The solid-state synthesis like the other two preparation techniques relies on diffusion processes between Y$_2$O$_3$ and SiO$_2$ precursors to form X1/X2-YSO. However, due to the lack of an aqueous phase to improve homogeneity and facilitate formation of precursor sizes, it is less likely to form at the synthesis temperatures used here. Greater temperatures, or the addition of a flux would likely lead to the synthesis of X2-YSO, however the flux may introduce unwanted optical defects. The advantage of using the sol-gel method is that it allows both the yttrium and silica sources to be closer than conventional solid-state methods, lowering the required synthesis temperature while simultaneously minimising Y$_2$O$_3$ impurities. For this advantage, in conjunction with the lack of other parasitic phases that form in solution combustion methods, the sol-gel method was selected for the growth of various lanthanide dopants for optical characterisation.

\begin{figure}[ht!]
	\centering
	\includegraphics[width=\linewidth]{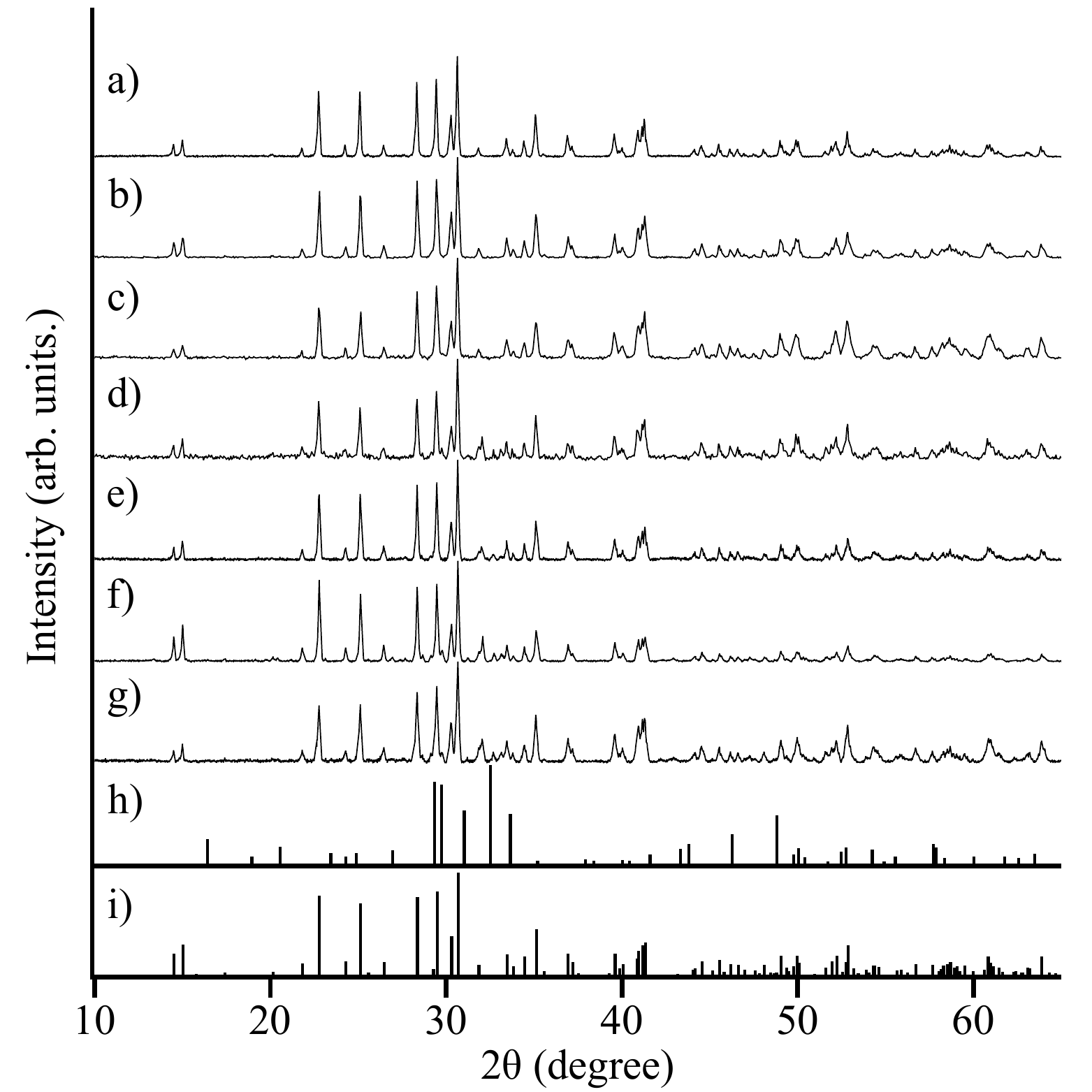}
	\caption{X-ray diffractograms of lanthanide-doped YSO micro-crystals. a) YSO:Ce$^{3+}$, b) YSO:Pr$^{3+}$, c) YSO:Nd$^{3+}$, d) YSO:Eu$^{3+}$, e) YSO:Ho$^{3+}$, f) YSO:Er$^{3+}$, g) YSO:Yb$^{3+}$ h) X1-YSO reference ICCD card No. 00-052-1810  i) X2-YSO reference ICCD card No. 01-084-4282 }
	\label{ysoxrd}
\end{figure}

\begin{figure}[ht!]
	\centering
	\includegraphics[width=\linewidth]{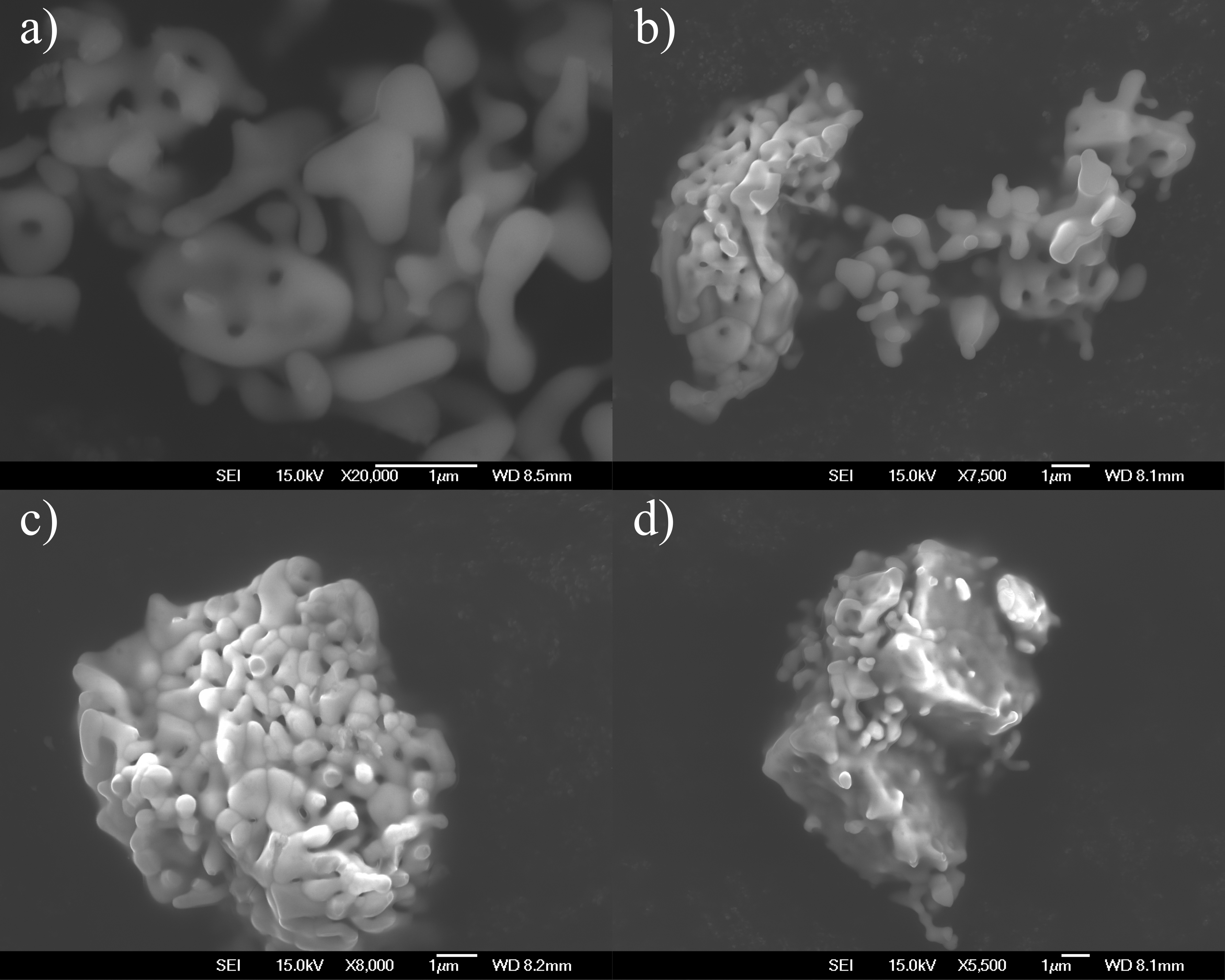}
	\caption{Representative SEM images of lanthanide-doped YSO showing a range of crystal morphologies present in all YSO samples prepared via the sol-gel route. a) Loose globular single YSO microcrystals b) semi-fused agglomerates c) fused agglomerates and d) monolithic YSO crystals}
	\label{ysosem}
\end{figure}

X-ray diffractograms for seven different, sol-gel prepared, rare-earth doped YSO samples are presented in Fig \ref{ysoxrd}. It can be seen from visual inspection that all samples are in good agreement with the reference data for X2 phase YSO (ICCD Card No. 01-084-4282), showing the reliability of this synthesis method to produce a wide range of YSO:Ln$^{3+}$ crystals. Additional diffraction peaks are noted at 32$^{\circ}$ for samples doped with heavier lanthanide ions (Eu - Yb) that are not present in the lighter doped samples (Ce - Nd). The additional features, while not present in the reference data, are predicted by the calculated diffraction patterns based on the YSO crystal structure. This may be due to the smaller lanthanide ions inducing less strain on the lattice, and, therefore, greater crystallinity. The inferred average crystallite size of the crystals was found to be 65 +/- 7\,nm. The SEM indicates that despite the small crystallite size, they are agglomerated and fused into larger structures. The most common structure is a 1-2\,$\mu$m structure (Fig \ref{ysosem}a) they appear to be fused structures of smaller sub-micron globules, connected via an elongated filament. These structures likely form during the high-temperature sintering with nucleated material from the precipitation step, only to begin fusing together with one another at elevated temperatures leading to globular morphology. However, it is noticeable that some of these structures have been cleaved off, indicating it is possible to miniaturise these materials further through ball milling or at least break up the agglomerates into their mostly sub-micron globules. These $\sim$1\,$\mu$m structures can also fuse into yet larger structures progressively, from moderately sized agglomerates (Fig \ref{ysosem}b,c) to more traditional monolithic crystal structures (Fig \ref{ysosem}d). It implies a heterogeneous nucleation process with some of the precursor crystals growing much larger than others. This distribution could also be caused by non-uniform packing during the sintering process, as well as uneven heat distribution leading to larger monoliths forming alongside the smaller globules. Despite the heterogeneous size, the optical linewidths (discussed in later sections) are relatively narrow, indicating the size distribution is not having an appreciable impact on the optical properties. 

\subsection{Optical Characterisation}
As stated in the introduction above, Y$_2$SiO$_5$ contains two substitutional Y$^{3+}$ sites, site 1 (sometimes denoted as Y$_1$) and site 2 (sometimes denoted as Y$_2$), each with C$_1$ point group symmetry and these are distinguished by their co-ordination numbers of six and seven, respectively. The relative populations of these two centres vary across the lanthanide series however in all materials, with the possible exception of Ce$^{3+}$ doping \cite{yashar1}, both centres are observed in non-site selective measurements such as infrared to optical absorption spectra. Below we present spectroscopic measurements of three selectively chosen materials.

\subsubsection{YSO:Nd$^{3+}$}

\begin{figure*}[ht!]
	\centering
	\includegraphics[width=\linewidth]{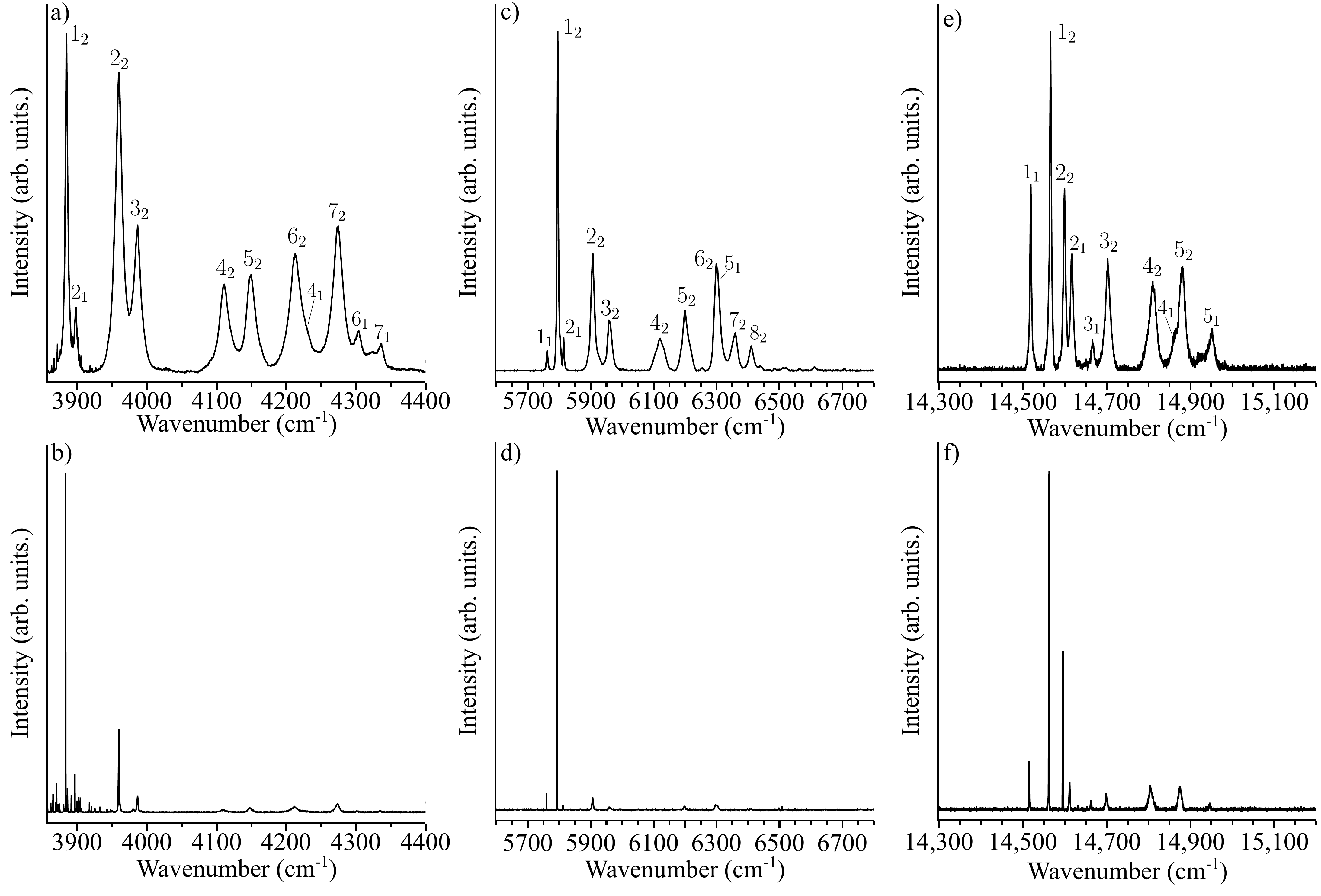}
	\caption{Representative absorption spectra of YSO:Nd$^{3+}$ recorded for samples cooled to 10\,K. The a) $^4$I$_{9/2}$$\longrightarrow$$^4$I$_{13/2}$,  c) $^4$I$_{9/2}$$\longrightarrow$$^4$I$_{15/2}$ and e) $^4$I$_{9/2}$$\longrightarrow$$^4$F$_{9/2}$ transitions measured for a pelletised microcrystal ensemble doped at 2 molar percent of Nd$^{3+}$. The b) $^4$I$_{9/2}$$\longrightarrow$$^4$I$_{13/2}$, d) $^4$I$_{9/2}$$\longrightarrow$$^4$I$_{15/2}$ and f) $^4$I$_{9/2}$$\longrightarrow$$^4$F$_{9/2}$ transitions measured for a bulk crystal doped at 0.02 molar percent of Nd$^{3+}$. The absorption features are labelled numerically to indicate the Stark sublevel upon which the transition terminates with a subscript indicating to which site the transition is assigned. Sharp spectral features observed for the $^4$I$_{13/2}$ multiplet in the 3850-3950 cm$^{-1}$ region are associated with residual atmospheric water absorption within the N$_{2}$ gas purged FTIR beam path.}
	\label{ysoNd}
\end{figure*}

\noindent\\
YSO:Nd$^{3+}$ has been investigated for quantum memory purposes, having shown a great deal of promise for AFC based quantum memories operating in the telecommunication band \cite{clausen2011quantum,usmani2010mapping}, as well as in proof of concept Purcell enhanced nanophotonic cavities \cite{zhong2015nanophotonic}. 

Representative 10~K absorption spectra of both a pelletised micro-crystal ensemble (having a 2 molar \% Nd$^{3+}$ doping level and a thickness of approximately 200~$\mu$m) and a bulk crystal (having a 0.02 molar \% doping level and a thickness marginally greater than 5~mm) are given in Fig \ref{ysoNd}, where the transition assignments are based on refs \cite{yashar3, yashar2}. In YSO crystals, the lighter lanthanide ions predominantly substitute into the 7-fold coordinated crystallographic site (site 2). Any changes to this site occupancy could be indicative of reduced crystallinity or different formation pathways. Forming the crystals through a sol-gel and sintering process could lead to slightly different site occupancy due to the lower temperatures used (1400\,$^{\circ}$C vs $>$2000\,$^{\circ}$C \cite{shoudu1999czochralski}) with less mobile ions unable to find an ideal, minimum energy configuration. However, as judged from Fig \ref{ysoNd} the relative site occupancy is observed to be very comparable to that of the bulk crystal. We note that the doping concentrations are nominal, determined from the ratios of precursors.

It is immediately clear from the absorption spectra that the optical linewidths for the bulk crystal are much lower than for the microcrystals. This is due to the much lower dopant concentration in the only available bulk crystal. It is also significantly easier to ensure good thermal contact of a bulk crystal to the copper cooling stage of a closed cycle compressor, thereby readily obtaining the 10~K base temperature. Any modest difference in the sample temperature aside, the broader absorption linewidths observed for the microcrystal ensemble can reasonably be attributed to solid-state strain induced by the 2 molar percent dopant concentration used; this having the effect of increasing inhomogeneous line broadening. The effect of such broadening is most evident in the lower energy transitions of a given multiplet. Leading to a change in the peak intensities between the bulk and micro-crystals.
The linewidths observed here are similar to those measured for YSO:Sm$^{3+}$ bulk crystals doped at 0.5 molar \% \cite{jobbitt2021laser}, suggesting minimal disorder has been induced into the crystal by the synthesis method itself. Apart from the obvious intensity changes due to the inhomogeneous linewidths when compared to the bulk crystal, there is good agreement between the bulk and micro-crystals. There is no indication of additional defect sites from either X1-YSO, any of the pyrosilicate phases e.g. $\beta$-Y$_2$SiO$_7$, or Y$_2$O$_3$; the additional weak transitions observable between $6500-6700$\,cm$^{-1}$ arise from unintentional Er$^{3+}$ impurities. 

\subsubsection{YSO:Eu$^{3+}$}

\noindent\\
YSO:Eu$^{3+}$ has been at the centre of attention for quantum memory applications \cite{jobez2014cavity,timoney2013single,timoney2012atomic}, due to the prospect of coherence times of anything up to $\sim$23 days \cite{konz2003temperature} (6 hours having been experimentally observed \cite{zhong2015optically}) and the fact that Eu$^{3+}$ is very conveniently optically addressable in the visible via its $^7$F$_0$$\longrightarrow$$^5$D$_0$ transition. 

\begin{figure}[ht!]
	\centering
	\includegraphics[width=\linewidth]{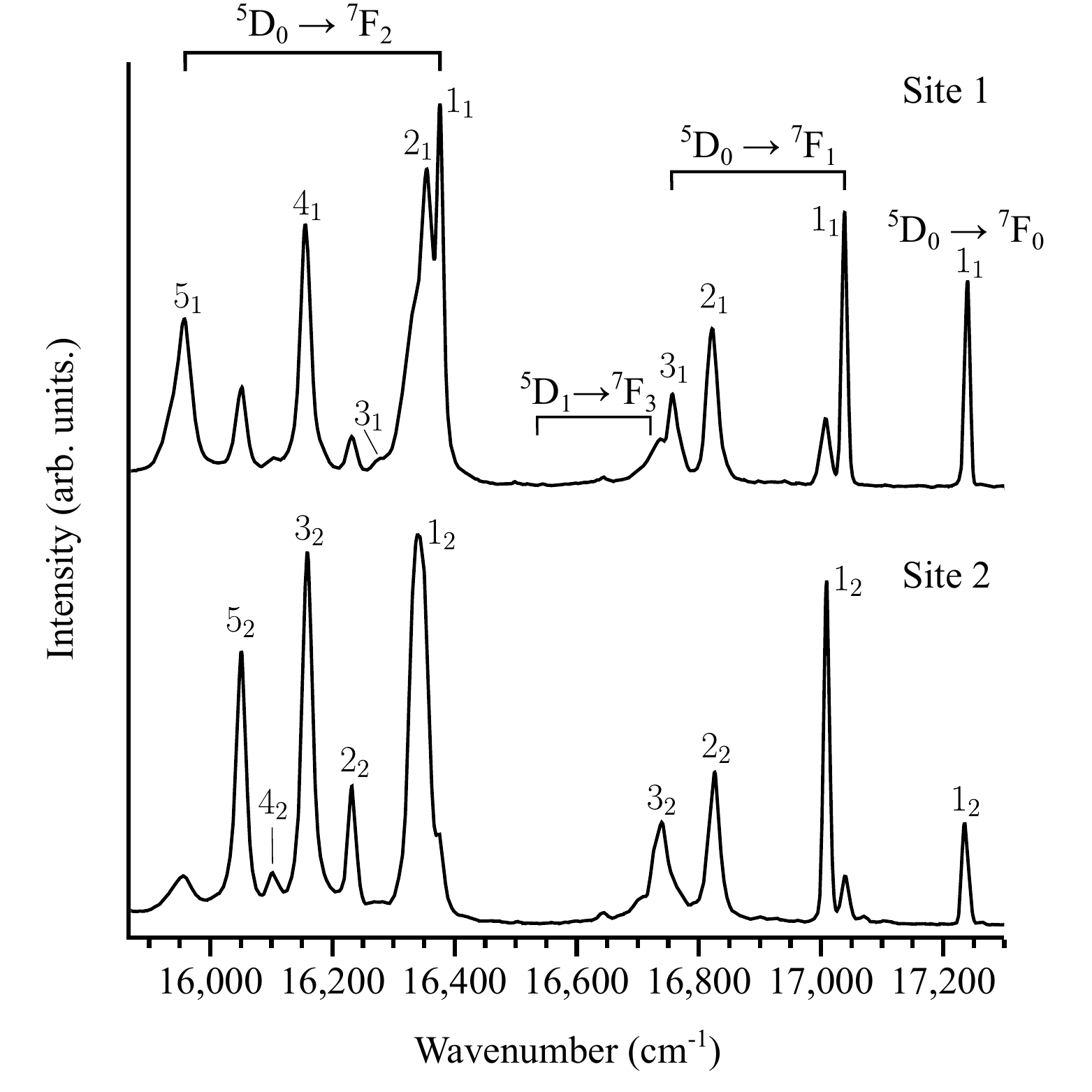}
	\caption{10\,K laser site-selective Eu$^{3+}$ fluorescence for site 1 and site 2 in YSO microcrystals. Site 1 was excited at 18952 cm$^{-1}$ whilst site 2 was excited at 18956 cm${-1}$. Fluorescence is observed from $^{5}$D$_{0}$ and $^{5}$D$_{1}$, with $^{5}$D$_{0}$ fluorescence labelled numerically according to the Stark sub-level of the terminating $^{7}$F term multiplet. The subscript indicates the relevant crystallographic site.}
	\label{ysoeu5d0}
\end{figure}

10K laser site-selective spectroscopy of the Eu$^{3+}$ ion was performed for both site 1 and site 2; employing excitation into both the $^5$D$_2$ (at 21544 cm$^{-1}$ for site 1 and at 21527 cm$^{-1}$ for site 2) and $^5$D$_1$ (at 18952 cm$^{-1}$ for site 1 and at 18956 cm$^{-1}$ for site 2) multiplets. True site selectivity could not be achieved. This is attributed to inter-site energy transfer processes which are known to occur in YSO crystals at modest dopant concentrations (at 0.5 molar \% or potentially lower) \cite{jobbitt2020energy}, noting that the Eu$^{3+}$ concentration used here was 2 molar \%. We observe that site 2 $\rightarrow$ site 1 energy transfer is more efficient than vice versa, however two way energy transfer between sites occurs unlike the observation for Sm$^{3+}$ in bulk YSO crystals \cite{jobbitt2020energy}.

\begin{figure}[ht!]
	\centering
	\includegraphics[width=\linewidth]{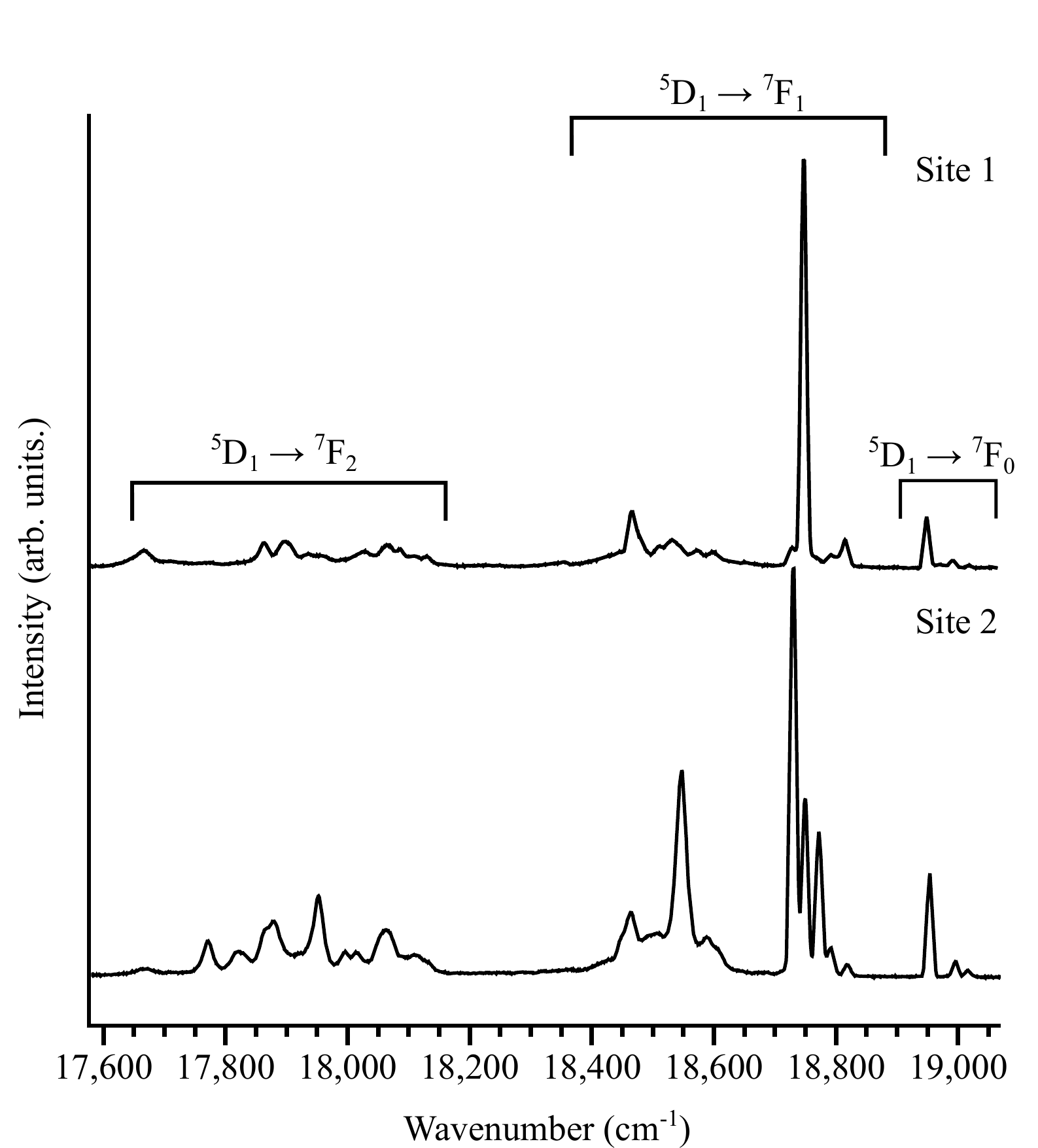}
	\caption{10\,K site-selective fluorescence of the $^5$D$_1$$\longrightarrow$$^7$F$_{0,1,2}$ transitions for YSO:2\%Eu$^{3+}$ microcrystals. Exciting the $^5$D$_2$ multiplet at 21544 cm$^{-1}$ for site 1 and at 21527 cm$^{-1}$ for site 2.}
	\label{ysoeu5d12}
\end{figure}

\begin{figure}[ht!]
	\centering
	\includegraphics[width=\linewidth]{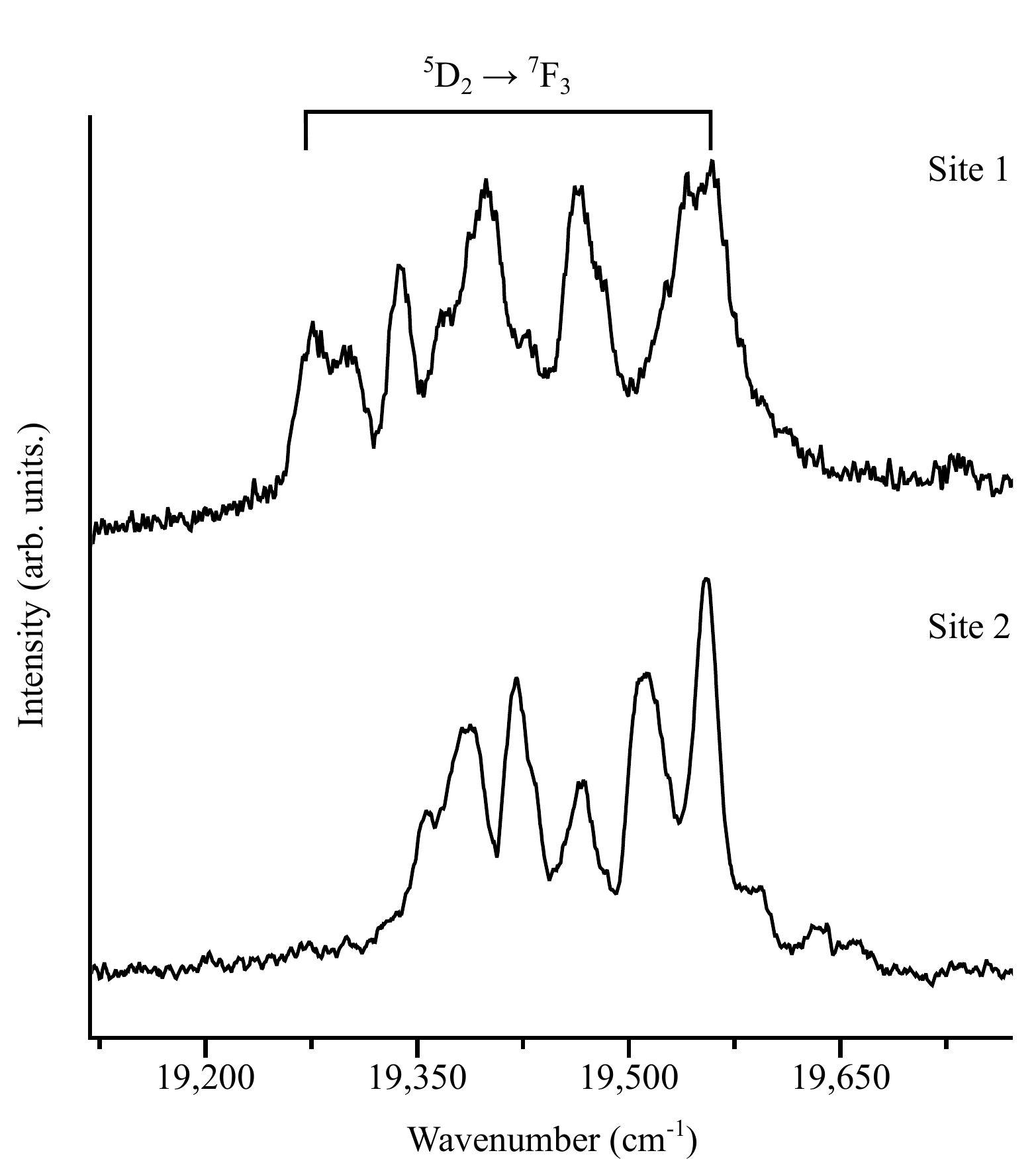}
	\caption{10\,K site-selective fluorescence of the $^5$D$_2$$\longrightarrow$$^7$F$_3$ transition for YSO:2\%Eu$^{3+}$ microcrystals. Exciting the $^5$D$_2$ multiplet at 21544 cm$^{-1}$ for site 1 and at 21527 cm$^{-1}$ for site 2. }
	\label{ysoeu5d2}
\end{figure}

Figure \ref{ysoeu5d0} shows site-selective fluorescence spectra, primarily from $^{5}$D$_{0}$, obtained for excitation of the $^5$D$_1$ multiplet. Overall the spectra are in excellent agreement with those obtained for the bulk crystal and as such, transition assignments are based on those of  K{\"o}nz \cite{konz2003temperature}. Figure \ref{ysoeu5d12} shows laser site-selective fluorescence primarily from $^{5}$D$_{1}$ with excitation into the $^{5}$D$_{2}$ multiplet. $^{5}$D$_{2}$ fluorescence is observable (Fig \ref{ysoeu5d2}), albeit very weak; apparently due to fast and efficient non-radiative relaxation. It is notable that hotlines can be observed in the $^{5}$D$_{1}\longrightarrow ^{7}$F$_{\rm J}$ fluorescence, perhaps most readily apparent for the $^{7}$F$_{0}$ or $^{7}$F$_{1}$ spectra. This suggests that a liquid helium immersion dewar would be a better option to truly cool pelletised micro- or nano-particles below 10~K.  

\begin{figure}[ht!]
	\centering
	\includegraphics[width=\linewidth]{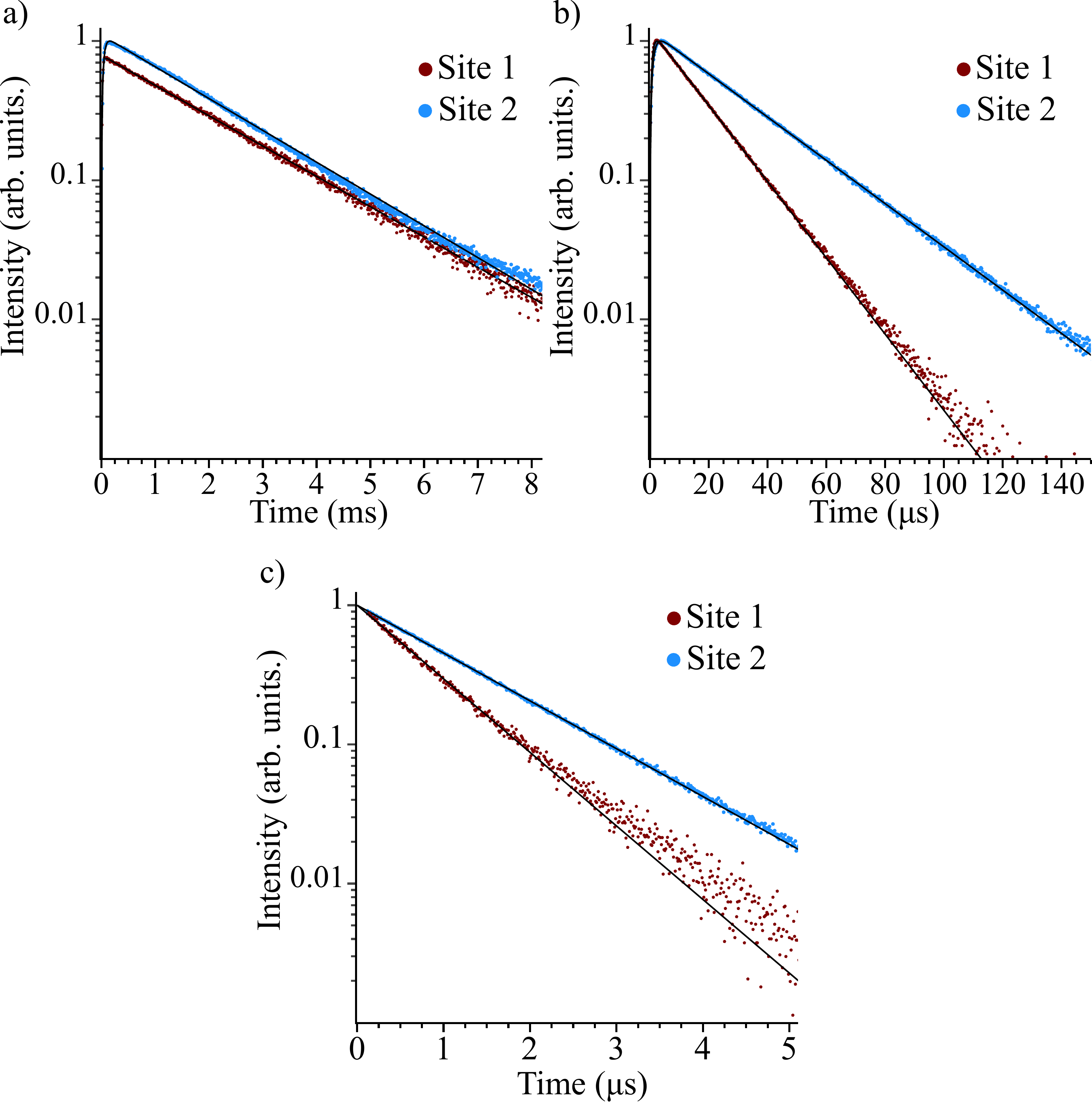}
	\caption{10\,K fluorescence transients for YSO:2\%Eu$^{3+}$ microcrystals. a) $^5$D$_{0}$, b) $^5$D$_{1}$ and c) $^5$D$_{2}$.}
	\label{yso_eulifetimes_plot}
\end{figure}

Site selective measurement of the $^{5}$D$_{0}$, $^{5}$D$_{1}$ and $^{5}$D$_{2}$ fluorescence lifetimes are presented in Fig \ref{yso_eulifetimes_plot}.
The observed temporal transient being for the most part, well-approximated by a single exponential decay, with decay constants as given in table \ref{yso_eu_lifetimes}. The short time behaviour of the $^{5}$D$_{1}$ and $^{5}$D$_{0}$ fluorescence is simply governed by a risetime reflecting higher frequency excitation. It is notable that the $^{5}$D$_{1}$ lifetimes are faster than those measured for the bulk crystal. The surface-to-volume ratios for these microcrystals are such that one would not expect surface states to play a significant quenching role, and in any case, such effects would influence the $^{5}$D$_{0}$ lifetime as well. As such, we speculate that the reduction in the $^{5}$D$_{1}$ lifetimes are a reflection of the higher Eu$^{3+}$ concentrations used for the microcrystals and thus cross-relaxation or inter-ionic energy transfer processes.

\begin{table}[ht!]
\centering
	\caption{10~K fluorescence decay times for YSO:2\%Eu$^{3+}$ microcrystals}
\begin{tabular}{ccccc}
\hline
\multirow{2}{*}{Multiplet} & \multicolumn{2}{c}{Site 1} & \multicolumn{2}{c}{Site 2} \\ \cline{2-5} 
                           & Micro        & Bulk        & Micro        & Bulk        \\ \hline
$^5$D$_0$                        & 1.99\,ms         & 1.94\,ms \cite{konz2003temperature}           &   1.90\,ms          &     1.67\,ms \cite{konz2003temperature}       \\
$^5$D$_1$                        &  15.8\,$\mu$s            &    32.0\,$\mu$s \cite{shen19947}         &    28.0\,$\mu$s          &    42.0\,$\mu$s \cite{shen19947}         \\
$^5$D$_2$                        &  0.8\,$\mu$s        &   -      &    1.3\,$\mu$s          &     -        \\ \hline
\end{tabular}
\label{yso_eu_lifetimes}
\end{table}
  

\begin{figure}[ht!]
	\centering
	\includegraphics[width=\linewidth]{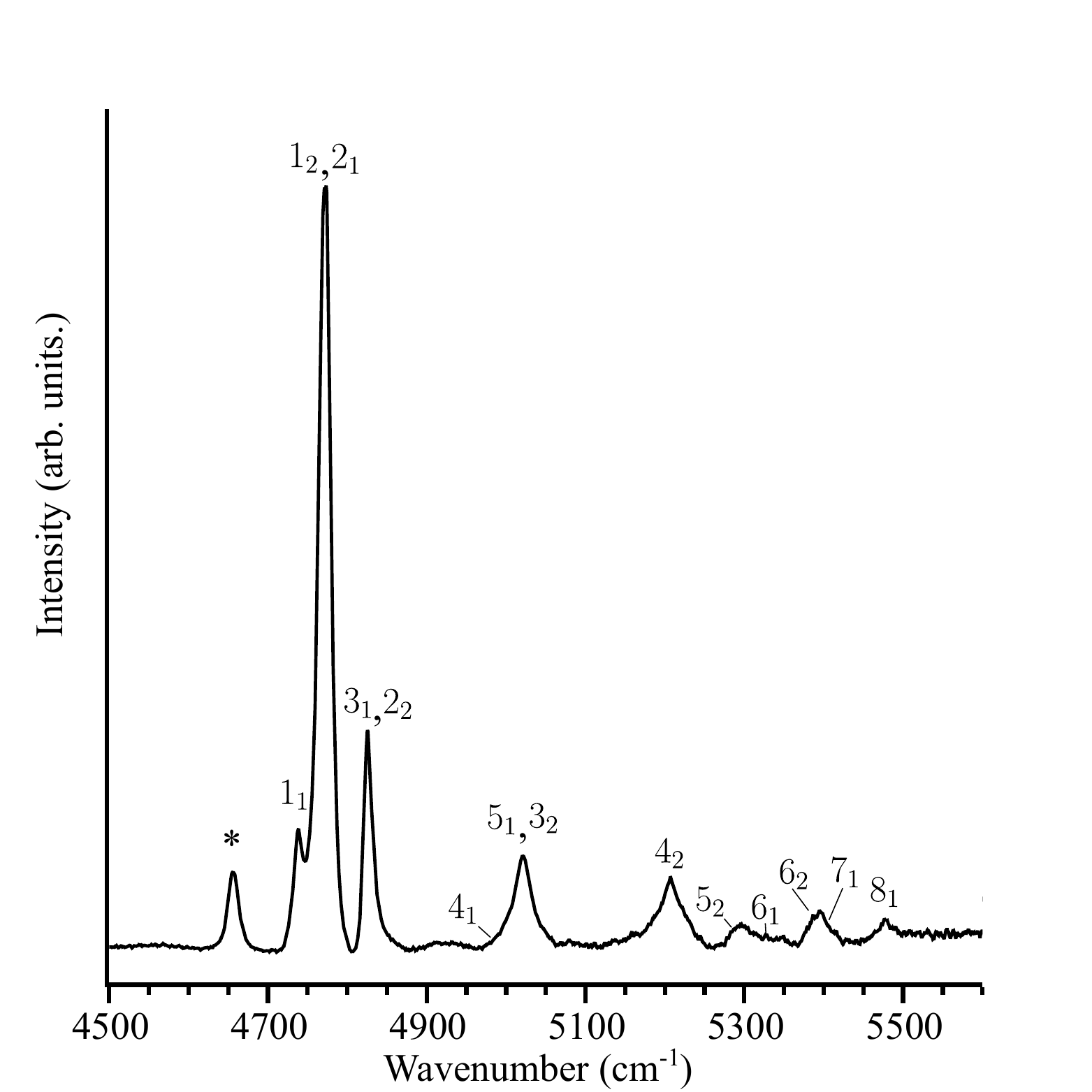}
	\caption{10\,K infrared absorption of the $^7$F$_6$ multiplet for a 200 $\mu$m thick pelletised YSO:2\%Eu$^{3+}$ microcrystal ensemble. The absorption features are labelled numerically to indicate the Stark sublevel upon which the transition terminates with a subscript indicating to which site the transition is assigned. * indicates an unassigned transition. }
	\label{ysoeu7F6}
\end{figure}
In our microcrystal experiments, fluorescence could be observed up to the $^{7}$F$_{3}$ multiplet. However, the $^7$F$_6$ multiplet was readily measurable in absorption as seen in Fig \ref{ysoeu7F6}, with the observed transitions giving good agreement with the assignments of K{\"o}nz, determined from fluorescence. As was the case for YSO:Nd$^{3+}$, no obvious defect centres can be observed.

\subsubsection{YSO:Er$^{3+}$}

\noindent\\
Erbium has a strong transition near 1.5\,$\mu$m, allowing erbium doped devices to integrate with existing telecommunications infrastructure, motivating the development of microwave to optical transducers using YSO:Er$^{3+}$ \cite{jevon2015}; a key ingredient in quantum hybrid signal processing with superconducting qubits. The narrow homogeneous linewidth of 50\,Hz \cite{sun2002recent} coupled with reasonably large hyperfine splittings, provide for significant bandwidth motivating work into this system \cite{lauritzen2010telecommunication,rakonjac2020long,horvath2019extending}.       

\begin{figure}[ht!]
	\centering
	\includegraphics[width = \linewidth]{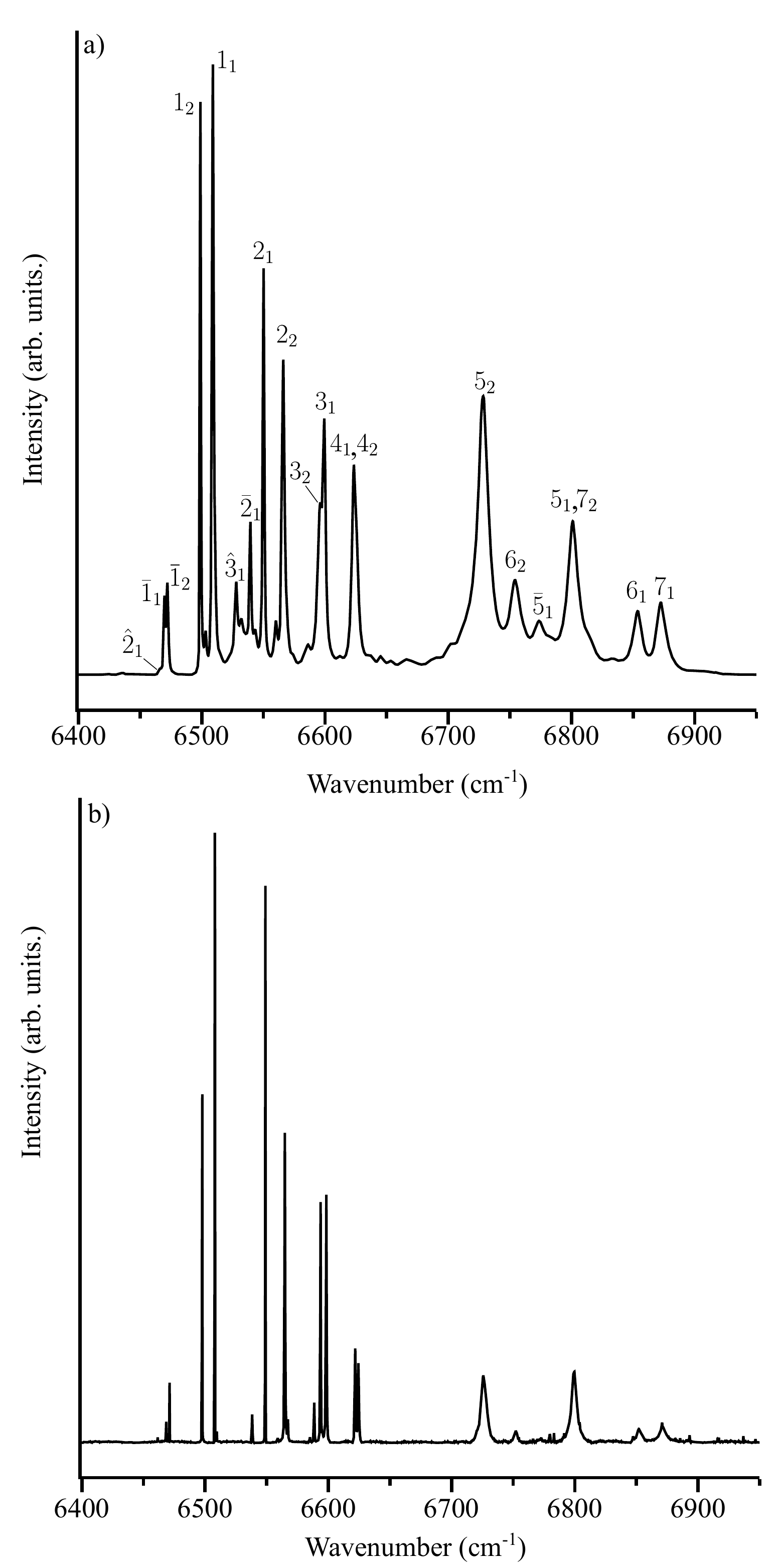}
	\caption{10\,K $^4$I$_{15/2}$$\longrightarrow$$^4$I$_{13/2}$ absorption of YSO:Er$^{3+}$; a) for a 200 $\mu$m thick pelletised microcrystal doped at 2 molar percent Er$^{3+}$ and b) for a 5 mm thick bulk crystal doped at 0.005 molar percent of Er$^{3+}$. The absorption features are labelled numerically to indicate the Stark sublevel upon which the transition terminates with a subscript indicating to which site the transition is assigned. An overbar indicates a thermally activated transition.}
	\label{yso_er}
\end{figure}

Representative YSO:Er$^{3+}$ infrared absorption spectra for the $^{4}$I$_{13/2}$ and $^{4}$I$_{11/2}$ multiplets, measured for both a 200 $\mu$m thick pelletised microcrystal ensemble doped at 2 molar percent and a 5 mm thick bulk crystal doped at 0.005 molar percent, both nominally at a sample temperature of 10~K are given in Fig \ref{yso_er}. As noted previously for Eu$^{3+}$, thermally activated transitions can be observed for the microcrystal samples. These transitions are not as strong in the spectra for the bulk crystal, again due to poorer thermal contact with the microcrystal samples. This aside, the measured spectra are comparable with the exception of linewidth and intensity differences, as was noted with the Nd$^{3+}$ samples.
In fact, the narrowest linewidth measured for the microcrystals in the 1.5\,$\mu$m region, is the site 2 Z$_1$ to Y$_1$ transition, with a linewidth of 0.43\,cm$^{-1}$ ($\sim$13 GHz); this inhomogeneous linewidth approaches the requirement of $<$10 GHz for efficient Purcell enhancement in nanophotonic cavities \cite{zhong2017nanophotonic}. Given the high dopant concentration of two molar percent, this result is promising. It implies a crystallinity comparable to the bulk crystal and that much better results could be obtained with a lower dopant concentration and potentially higher resolution spectroscopic measurements. 

\begin{figure}[ht!]
	\centering
	\includegraphics[width = \linewidth]{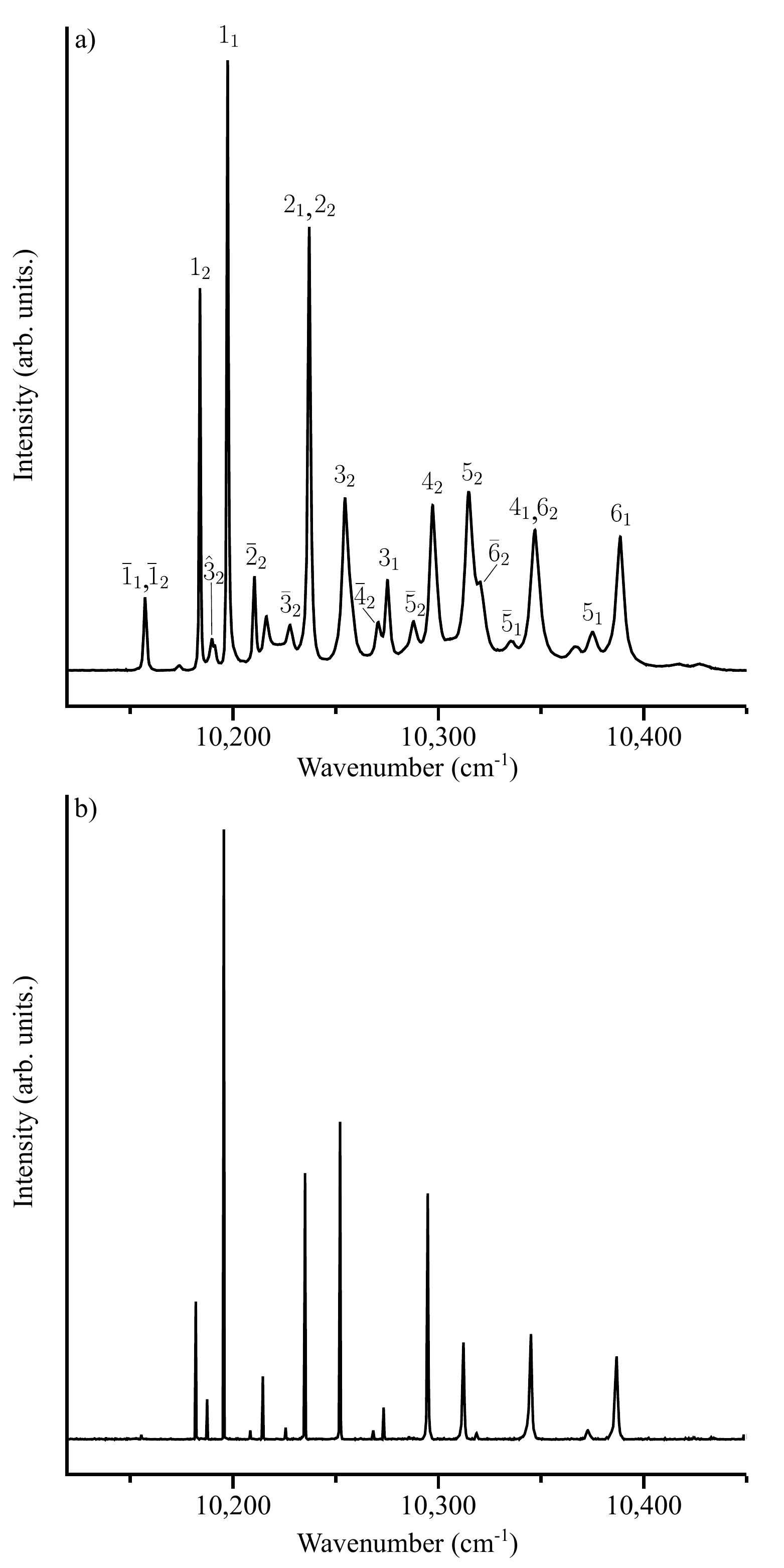}
	\caption{10\,K $^4$I$_{15/2}$$\longrightarrow$$^4$I$_{11/2}$ absorption of YSO:Er$^{3+}$; a) for a 200 $\mu$m thick pelletised microcrystal doped at 2 molar percent Er$^{3+}$ and b) for a 5 mm thick bulk crystal doped at 0.005 molar percent of Er$^{3+}$. The absorption features are labelled numerically to indicate the Stark sublevel upon which the transition terminates with a subscript indicating to which site the transition is assigned. An overbar indicates a thermally activated transition.}
	\label{yso_er2}
\end{figure}

\section{Conclusion}
We have prepared lanthanide doped Y$_{2}$SiO$_{5}$ microcrystals using three different techniques; solution combustion, solid state and sol-gel synthesis. Whilst all three methods produce X2 phase Y$_{2}$SiO$_{5}$, the solution and solid state synthesis methods also yield unacceptably high quantities of Y$_{2}$O$_{3}$ and $\beta -$Y$_{2}$Si$_{2}$O$_{7}$, for the growth temperatures used here. As such, we conclude that sol-gel synthesis is the most appropriate method to produce high-quality X2 phase Y$_{2}$SiO$_{5}$ microcrystals; nanocrystals being readily obtainable through the addition of a ball milling procedure.

A selective spectroscopic assay of Nd$^{3+}$, Eu$^{3+}$ and Er$^{3+}$ doped Y$_{2}$SiO$_{5}$ microcrystals indicates that the as-prepared samples are of high optical quality. Inhomogeneous linewidths as low as 12 GHz are measured from 10\,K absorption experiments, whilst fluorescence lifetimes are inferred to be of the same order of magnitude as the bulk crystals. Inter-site energy transfer is observed in the microcrystal systems at the two molar percent dopant levels used in this work. The narrow inhomogeneous linewidths measured in particular for YSO:Er$^{3+}$ are promising for direct integration into future quantum information processing devices.

\clearpage

\bibliographystyle{elsarticle-num}
\bibliography{bib}


\end{document}